\documentclass[11pt]{article}
\usepackage{amsmath}
\usepackage{amssymb}
\usepackage{bm}
\usepackage{graphicx}
\usepackage{makecell}
\usepackage{url}

\DeclareMathOperator{\sgn}{sgn}%
\DeclareMathOperator{\sat}{sat}%

\begin{document}

\title{Sliding Mode Control of Cardiac Rhythms in the Sinoatrial Node using Gaussian Process Regression}

\author{
Gabriel da Silva Lima\\
University of Turku
\and
Marcelo Amorim Savi\\
Universidade Federal do Rio de Janeiro
\and
Wallace Moreira Bessa\\
University of Turku
}

\maketitle

\abstract{The Sinoatrial node (SA), also called natural pacemaker, is responsible to initiate the heart electrical activity, usually represented by electrocardiograms (ECGs). Abnormalities at the SA node can produce disordered heart rhythms or, in other words, cardiac arrhythmia that are visualized in the ECGs. The development of control strategies to stabilize the cardiac rhythm at the natural pacemaker can provide efficient ways to deal with and avoid some heart pathology. This paper investigates the use of a robust controller based on sliding modes for cardiac rhythms at the SA node in order to induce normal rhythms from pathological responses. Embedded into this controller, a Gaussian process regressor is utilized to predict and compensate modeling uncertainties and disturbances. A mathematical model that presents close agreement with experimental measurements is employed to represent the heart functioning. The adopted model comprises a network of oscillators formed by sinoatrial node, atrioventricular node (AV) and His-Purkinje complex (HP). Three nonlinear oscillators are employed to represent each one of the nodes that are connected by delayed couplings. The boudedness and convergence properties are investigated with a Lyapunov-like stability analysis. In order to evaluate the ability of the control law to deal with interpatient variability, the heart model is assumed to be not available to the controller designer, being used only in the simulator to assess the control performance. The results show that, by applying the proposed control scheme, abnormal rhythms can be avoided, turning the ECG closer to the expected normal behavior and preventing critical cardiac responses.}


\section{Introduction}\label{sec1}

\begin{figure}[htb]
    \centering
    \includegraphics[width=0.8\textwidth]{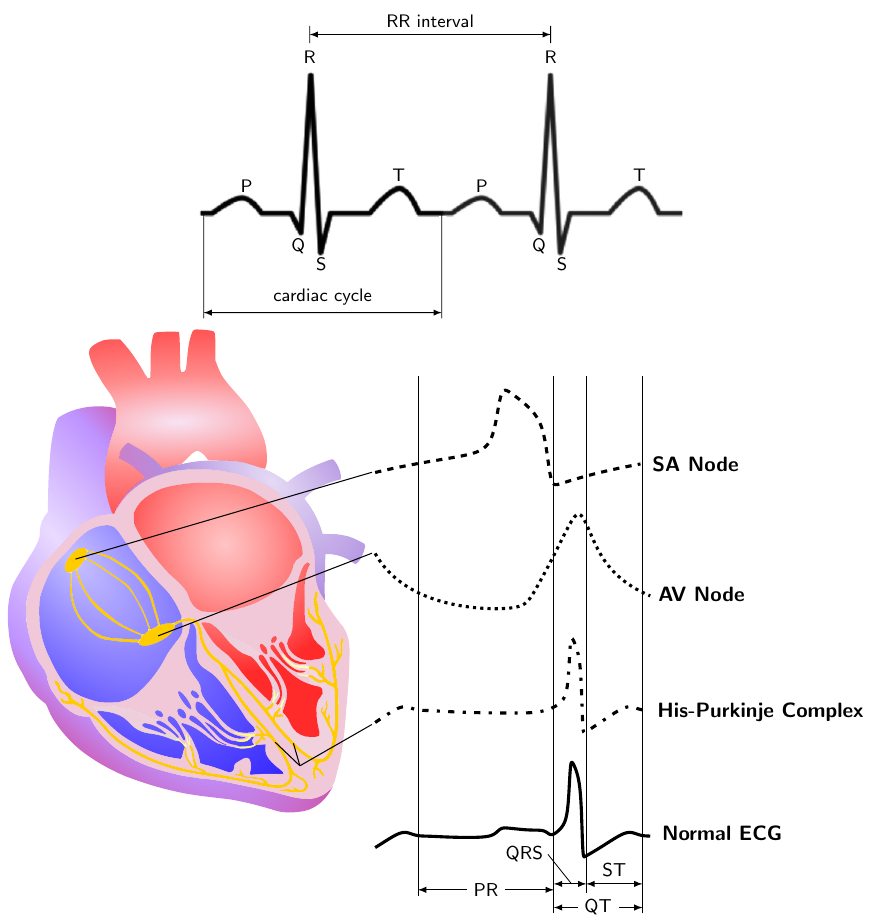}
    \caption{Schematic view of the heart, including the distinct waveforms for the corresponding specialized cells, and a normal ECG.}
    \label{fig:heart}
\end{figure}

One of the most widespread representations of the electrical activity of the heart is the electrocardiogram (ECG) due to not being required instruments introduction into the body. With an ECG in hand, the regularity and cardiac frequency can be analyzed as well as is possible to observe if some heart disorders are occurring \cite{shekatkar2017detecting}. The cardiac behavior is represented by electrical impulses captured as waves from different areas of the heart, as shown in the Figure~\ref{fig:heart} for a normal cycle. Its physiological dynamics operates as a combination of complicated self-excitatory elements, with the initial pulse starting in the sinoatrial (SA) node, the called natural pacemaker, and propagating to the atrioventricular (AV) node, reaching the bundle of His and, afterward, the Purkinje fibers, the HP complex. The ventricles contraction is provoked by the stimulus distributed by the Purkinje fibers in the myocardial cells. From a normal ECG, Figure~\ref{fig:heart}, it is important to call attention to three special features: the P wave, QRS complex and T wave. The first one is related to the impulse in the SA node. The QRS complex is consequence of the ventricular contraction. And the T wave represents the ventricular repolarization, i.e., the returning movement to the state in which the cardiac cells react to another stimulus.

In order to describe the heart dynamics, mathematical models have been widely used for different purposes. In this regard, it is important to mention the pioneer work of Van der Pol and Van der Mark \cite{van1928lxxii}. Grudzi{\'n}ski and {\.Z}ebrowski \cite{grudzinski2004modeling} proposed a modified Van der Pol (VdP) oscillator capable of properly presenting a description of the natural pacemaker. A model considering two asymmetrically coupled modified VdP oscillators to the SA and AV nodes was proposed by Dos Santos et al.\ \cite{dos2004rhythm}. Gois and Savi \cite{gois2009analysis} introduced a new oscillator to the previous model for the HP complex in order to represent the ECG signals. Each oscillator is based on the model due to Grudzi{\'n}ski and {\.Z}ebrowski and the system has bidirectional and asymmetric time-delayed couplings to represent the time spent on impulse transmissions. Cheffer et al.\ \cite{cheffer2021heart} improved the three-coupled oscillator model considering different coupling terms. Non-deterministic aspects were incorporated by considering random connections among oscillators \cite{cheffer2020random, cheffer2021analysis, cheffer2021uncertainty} in order to emulate some pathological dynamics. Another way is by means of external stimulation in the natural pacemaker \cite{cheffer2021biochaos}. 

Considering the challenge regarding to avoid cardiac arrhythmia, the control of the heart dynamics has been investigated by means of different approaches in order to analyse its potential application in rhythm management devices, such as artificial pacemakers and implantable defibrillators. The early studies about cardiac rhythm control were presented by Garfinkel et al.\ \cite{garfinkel1992controlling, garfinkel1995chaos} applying OGY method \cite{PhysRevLett.64.1196} on rabbit heart. Ferreira et al.\ \cite{ferreira2011chaos} employed time-delayed feedback control for natural pacemaker using a model proposed by reference \cite{gois2009analysis}. Afterward, Ferreira et al.\ \cite{ferreira2014chaos} employed the same technique for ECG signals built with a three-coupled oscillators. Lounis et al.\ \cite{lounis2020implementing} applied a high-order control method to the model proposed by Quiroz-Juarez et al.\ \cite{quiroz2019generation}, considering the stabilization of a desired unstable periodic orbit (UPO). Feedback linearization with state observers has also been applied by Gharesi et al.\ \cite{gharesi2021extended} to stabilize a cardiac system represented by an ECG built with a heterogeneous coupled oscillator model \cite{ryzhii2014heterogeneous}, converting a sinus tachycardia condition into a normal signal. Khan and Nigar \cite{khan2020combination} proposed a Lyapunov-based active controller considering the combination of projective synchronization in fractional-order chaotic system with disturbance and uncertainty. 

A suitable controller must deal not only with all nonlinearities inherent in the cardiac system but also with modeling inaccuracies and external disturbances. State observers can handle the first task \cite{gharesi2021extended} but may not be a suitable choice for the other issues. In turn, the intelligent control has already demonstrated be attractive to deal with uncertain nonlinear systems \cite{tanaka2013feedback, karar2018robust, deodato2019intelligent, xu2021finite, lima2021intelligent}. In this context, the sliding mode control (SMC) comes as an interesting controller framework because can deal with modeling uncertainties as well as can guarantee robustness to the system \cite{slotine1991applied}. However, the conventional SMC has a known drawback related to the chattering effect in the neighborhood of the sliding surface caused by noncontinuous switching terms. Although a thin boundary layer around the sliding surface can be adopted to smooth the high frequency vibration, this method also turns the perfect tracking into a tracking with guaranteed precision problem \cite{bessa2009some}. Therefore, the adoption of proper compensation functions is an approach used to reduce, or even eliminate, the steady-state control error of smooth controllers \cite{bessa2019adaptive, lima2020sliding, lima2021intelligent, da2022sliding}. 

Gaussian process regression (GPR) can serve as a suitable compensator providing not only a predicted value but also a distribution for the prediction of the unknown system dynamics as well as parametric inaccuracies. The Gaussian process can be understood as an extension of the Gaussian random variable of distributions over a function space \cite{williams2006gaussian}. The GPR can be adopted as a nonparametric model to represent unknown functions and to estimate both structured and unstructured uncertainties. For instance, the GPR has been combined with linear quadratic regulators (LQR) \cite{marco2016automatic, marco2017design} and active disturbance rejection control (ARDC) \cite{neumann2019data} in order to find the open parameters of controllers using experimental data. The GPR is also used to compensate model uncertainties in model predictive controllers (MPC) \cite{kocijan2004gaussian, cao2017gaussian, li2022learning, li2022online}. 

This paper proposes a sliding mode controller combined with a Gaussian process regressor to be applied in the natural pacemaker in order to regularize the ECG signal. The cardiac dynamics is modeled by a three-oscillator model \cite{cheffer2021heart} with the pathological behavior generated by an external excitation in the SA node \cite{cheffer2021biochaos}. The GPR scheme allows the controller to compensate the unknown cardiac dynamics and also provides an estimate of the uncertainty dispersion \cite{west2021use} which benefits the controller's robustness. The association between SMC and GPR has been proposed by Aran and Unel for a diesel engine control \cite{aran2018gaussian} using an offline training procedure. The algorithm proposed here is an online scheme to generate the training set necessary for the GP distribution allowing a continuously adaptation of the heart system on a supervised learning strategy called overlapping rolling windows. The boundedness and convergence properties of the control error are proven by means of the Lyapunov stability theory. Simulation results demonstrate that it is possible to regularize the heart signal from the natural pacemaker. Aiming a real application of the proposed controller, a basic circuit for an artificial pacemaker is presented in order to provide a more realistic scale of the control effort level.

\section{Mathematical Modeling}\label{sec2}
 
\begin{figure}[htb]
\centering
\includegraphics[height=0.25\textwidth]{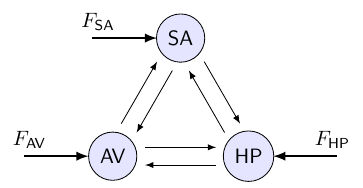}
\caption{Conceptual model of the cardiac system.}
\label{fig:concept_model}
\end{figure}

The electrical activity of the cardiac system can be modeled from the coupling of three nonlinear oscillators representing SA node, AV node and HP complex \cite{gois2009analysis, cheffer2021heart}, as shown in the conceptual model in Figure~\ref{fig:concept_model}. 

Asymmetrical and bidirectional connections are employed in order to build a general model that is capable to reproduce the electrical activity of the heart including normal and pathological functioning. The connections use time-delayed terms to represent the transmitting time spent among each one of the oscillators. Each oscillator is modeled by a modified van der Pol oscillator \cite{grudzinski2004modeling} since its dynamic response presents typical characteristics as limit cycle, synchronization and chaos \cite{gois2009analysis}. Under these assumptions, central nervous system stimuli are represented by self-excitatory behavior which means that external stimulus refers to situations different of the normal functioning and the cardiac system is governed by the following equations \cite{cheffer2021analysis}:

\begin{align}
\label{eq:cheffer_savi1}
\ddot{u}_{\text{SA}} = & \: F_{\text{SA}}(t) - \alpha_{\text{SA}}\,\dot{u}_{\text{SA}}(u_{\text{SA}} - \nu_{\text{SA}_1})(u_{\text{SA}} - \nu_{\text{SA}_2})+\nonumber\\
&-\frac{ u_{\text{SA}}(u_{\text{SA}} + d_{\text{SA}})(u_{\text{SA}} + e_{\text{SA}})}{d_{\text{SA}}\,e_{\text{SA}}} - k_{\text{\text{AV}}-\text{SA}}\,u_{\text{SA}} + k^{\tau}_{\text{AV}-\text{SA}}\,u_{\text{AV}}^{\tau_{\text{AV}-\text{SA}}}+\nonumber\\
&- k_{\text{HP}-\text{SA}}\,u_{\text{SA}} + k^{\tau}_{\text{HP}-\text{SA}}\, u_{\text{HP}}^{\tau_{\text{HP}-\text{SA}}} \\
\label{eq:cheffer_savi2}
\ddot{u}_{\text{AV}} = & \: F_{\text{\text{AV}}}(t) - \alpha_{\text{AV}}\,\dot{u}_{\text{AV}}(u_{\text{AV}} - \nu_{\text{AV}_1})(u_{\text{AV}} - \nu_{\text{AV}_2})+\nonumber\\
& - \frac{u_{\text{AV}} (u_{\text{AV}}  + d_{\text{AV}})(u_{\text{AV}}  + e_{AV})}{d_{\text{AV}}\,e_{\text{AV}}} - k_{\text{SA}-\text{AV}}\,u_{\text{AV}}  + k^{\tau}_{\text{SA}-\text{AV}}\,u_{\text{SA}}^{\tau_{\text{SA}-\text{AV}}}+\nonumber\\
&- k_{\text{HP}-\text{AV}}\,u_{\text{AV}} + k^{\tau}_{\text{HP}-\text{AV}}\,u_{\text{HP}}^{\tau_{\text{HP}-\text{AV}}} \\
\label{eq:cheffer_savi3}
\ddot{u}_{\text{HP}} = & \: F_{\text{HP}}(t) - \alpha_{\text{HP}}\,\dot{u}_{\text{HP}}(u_{\text{HP}} - \nu_{\text{HP}_1})(u_{\text{HP}} - \nu_{\text{HP}_2})+\nonumber\\
& -\frac{u_{\text{HP}}(u_{\text{HP}} + d_{\text{HP}})(u_{\text{HP}} + e_{\text{HP}})}{d_{\text{HP}}\,e_{\text{HP}}}- k_{\text{\text{SA}}-\text{HP}}\,u_{\text{HP}} + k^{\tau}_{\text{SA}-\text{HP}}\,u_{\text{SA}}^{\tau_{\text{SA}-\text{HP}}}+\nonumber\\
&- k_{\text{AV}-\text{HP}}\,u_{\text{HP}} + k^{\tau}_{\text{AV}-\text{HP}}\,u_{\text{AV}}^{\tau_{\text{AV}-\text{HP}}} 
\end{align}

Noting that indexes $m$ and $n$ represent SA, AV or HP, with $m \neq n$, equation terms and coefficients can be explained as follows: $k_{m-n}$ and $k_{m-n}^{\tau}$ are coupling coefficients between $m$ and $n$ nodes; $x_{i}^{\tau_{m-n}} = x_{i} (t-\tau_{m-n})$ are delayed terms, where $ \tau_{m-n} $ is the time delay. 

Although oscillator descriptions do not present spatial aspects, it can capture macroscopic spatial influences. On this basis, $F_{m}(t) = \rho_{m} \sin(\omega_{m} t)$ is an external excitation that has origin in spatiotemporal stimulus and therefore, it is considered as a reduced-order representation of spatiotemporal aspects. The harmonic form is motivated by mechanisms of atrial fibrillation (AF), which are represented by periodic behavior \cite{skanes1998spatiotemporal, jalife1998mechanisms}. Note that this external stimulus increases the system dimension by introducing an explicit time dependence based on spatiotemporal information.

The ECG can be represented by incorporating the signals of the three oscillators, being expressed as a linear combination of the state variables \cite{gois2009analysis}:
\begin{equation}
x = \text{ECG} = \beta_0 + \beta_1\,u_{\text{SA}} + \beta_2\,u_{\text{AV}} + \beta_3\,u_{\text{HP}}
\label{eq:ECG}
\end{equation}
\noindent with $ \beta_0 $, $ \beta_1 $, $ \beta_2 $ and $ \beta_3 $ being parameters, so that the derivative of the ECG with respect to $t$ becomes
\begin{equation}
\dot{x} =\frac{d}{dt}(\text{ECG}) =\beta_1\,\dot{u}_{\text{SA}} + \beta_2\,\dot{u}_{\text{AV}} + \beta_3\,\dot{u}_{\text{HP}}
\label{eq:dECG}
\end{equation}

Equations~\eqref{eq:ECG} and~\eqref{eq:dECG} can be used to represent the ECG phase space, favoring a qualitative assessment of cardiac cycle.

Since governing equations are presented in dimensionless form, it is interesting to define a dimensional time $\bar{t}$[s]$ : \bar{t} = \beta_t t$, where $\beta_t$ can be estimated by the ratio between real RR interval, RR$_\text{exp}$, and numerical RR interval, RR$_\text{num}$, $\beta_t = \text{mean}(\text{RR}_\text{exp}) / \text{mean}(\text{RR}_\text{num})$.

\subsection{Cardiac rhythms}

\begin{figure}[htb]
    \centering
    \includegraphics[width=\textwidth]{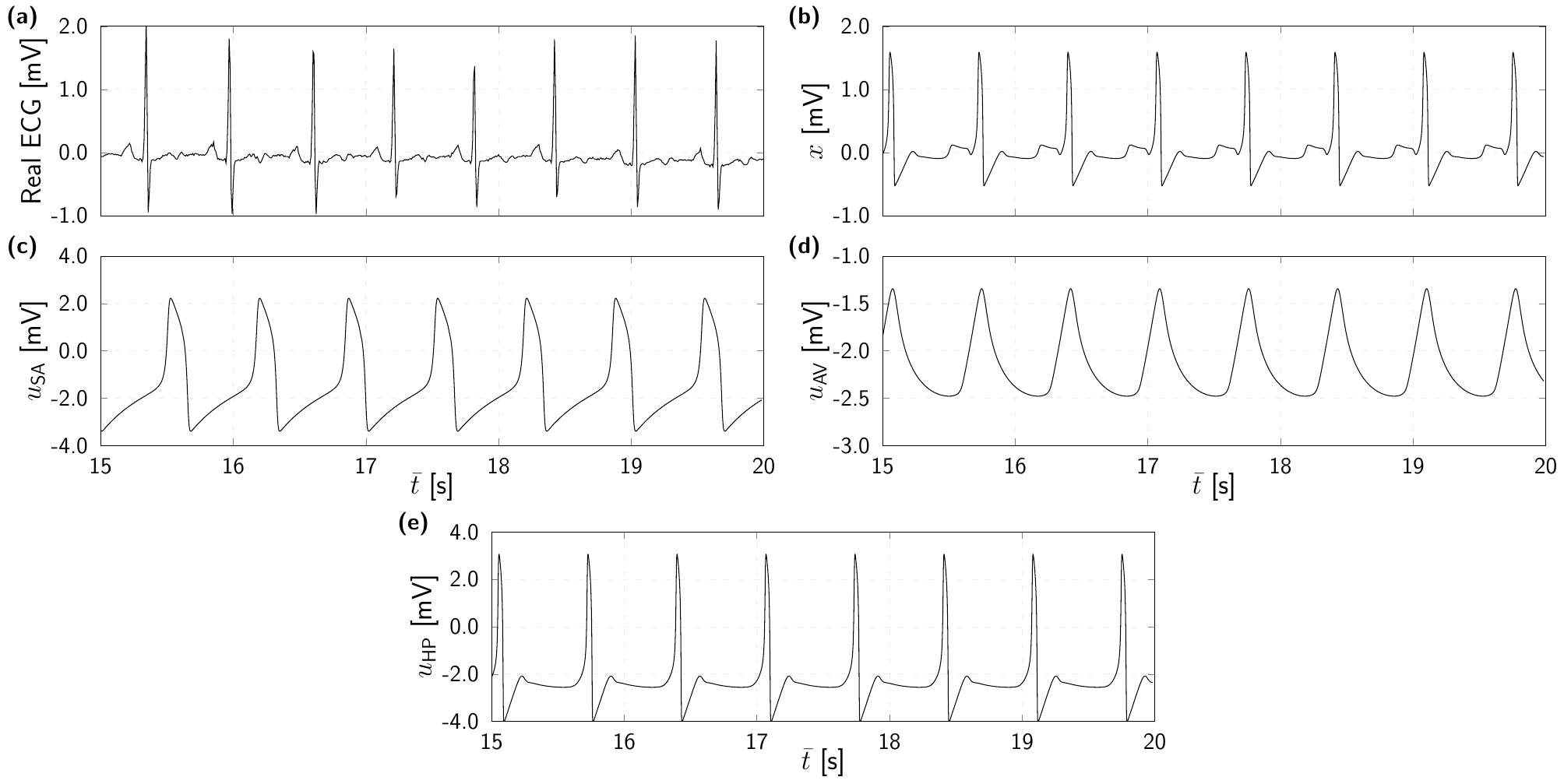}
    \caption{Normal cardiac rhythm: (a) real ECG signal \cite{PhysioNet}; (b) simulated time series; (c) SA, (d) AV, and (e) HP components of the simulated ECG.}
    \label{fig:normal}
\end{figure}

\begin{figure}[htb]
    \centering
    \includegraphics[width=\textwidth]{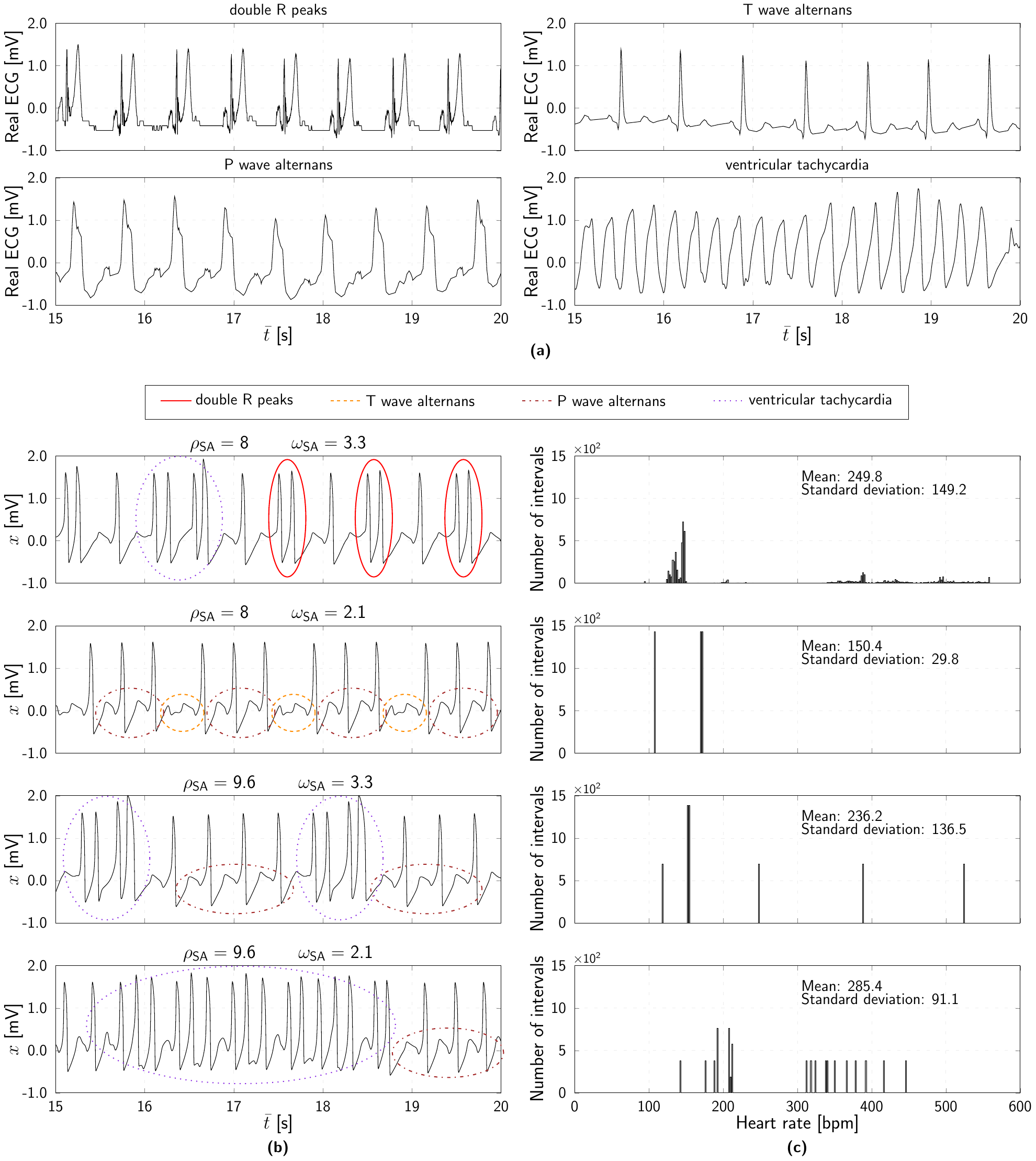}
    \caption{Pathological cardiac rhythm: (a) real ECG signals \cite{PhysioNet}; (b) simulated ECG signals; (d) histogram of the heart rate for each external stimulus.}
    \label{fig:diseases}
\end{figure}

In order to assess the model's ability to represent cardiac dynamics, first the normal rhythm is investigated. After this previous study and considering the importance of the SA node as natural pacemaker and, consequently, as the responsible by the initial excitement of the heart, the parameters of the external excitation $F_{\text{SA}}(t)$ are modified in order to identify some pathological behaviors. The dynamic model is numerically implemented in C++ using the fourth order Runge-Kutta method with sampling rate of 1 kHz. The model parameters are presented in the Table~\ref{tab:parameters}.  The numerical results obtained with the adopted model are compared with the real ECG data and are shown in Figures~\ref{fig:normal}-\ref{fig:diseases}, with the real data provided by the PhysioNet Databases \cite{PhysioNet}. In order to obtain a close agreement between the experimental and simulated data, the dimensional time is scaled to reproduce a normal ECG with heart rate in approximately 90 beats per minute (bpm), which corresponds to $\beta_t = 0.1048$. For all simulations, it is considered that $\beta_0 = 1$~mV, $\beta_1 = 0.06$, $\beta_2 = 0.1$, and $\beta_3 = 0.3$. Initial conditions are defined as $\bm{u}_0 = [-0.1, -0.6, -3.3 ]^\top$ and $\dot{\bm{u}}_0 = [0.025, 0.1, 2/3 ]^\top$, with $\bm{u} = [u_\text{SA} ~ u_\text{AV} ~ u_\text{HP}]^\top$. 

\begin{table}[htb]
\caption{Cardiac system parameters.}\label{tab:parameters}
\begin{tabular}{cccccccccc}
\hline
 \multicolumn{2}{c}{SA node} & \multicolumn{2}{c}{AV node} & \multicolumn{2}{c}{HP node} & \multicolumn{2}{c}{Couplings} & \multicolumn{2}{c}{Time delays} \\ \hline
$\alpha_{\text{SA}}$ & 3 & $\alpha_{\text{AV}}$ & 3 & $\alpha_{\text{HP}}$ & 7 & $k_{\text{SA-AV}}$ & 3 & $\tau_{\text{SA-AV}}$ & 0.8\\
$\nu_{\text{SA}_1}$ & 1 & $\nu_{\text{AV}_1}$ & 0.5 & $\nu_{\text{HP}_1}$ & 1.65 & $k_{\text{AV-HP}}$ & 55 & $\tau_{\text{AV-HP}}$ & 0.1 \\
$\nu_{\text{SA}_2}$ & -1.9 & $\nu_{\text{AV}_2}$ & -0.5 & $\nu_{\text{HP}_2}$ & -2 & $k_{\text{SA-AV}}^\tau$ & 3 \\
$d_{\text{SA}}$ & 1.9 & $d_{\text{AV}}$ & 4 & $d_{\text{HP}}$ & 7 & $k_{\text{AV-HP}}^\tau$ & 55 \\
$e_{\text{SA}}$ & 0.55 & $e_{\text{AV}}$ & 0.67 & $e_{\text{HP}}$ & 0.67 \\ \hline
\end{tabular}
\\
Parameters that are null for all cases are omitted.
\end{table}

\begin{figure}[htb]
\centering
\includegraphics[height=0.25\textwidth]{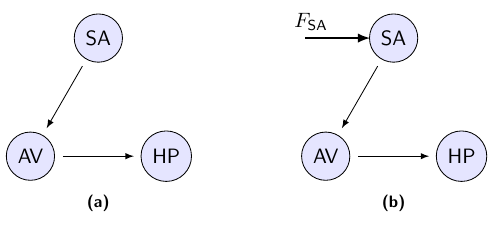}
\caption{Conceptual model of the real cardiac system without (a) and with (b) external excitation in the SA node.}
\label{fig:concept_model_real}
\end{figure}

Although the general representation considers the bidirectional coupling among all the three nodes, a real heart functioning is considered with the initial excitement starting in the SA node and being propagated to the AV node and reaching the HP node. For this reason, only the coupling terms related this unidirectional transmission is adopted in the implemented model. The conceptual model regarding this particular behavior is illustrated in the Figure~\ref{fig:concept_model_real}(a). The conceptual model \ref{fig:concept_model_real}(b) refers to the pathological behavior generated by external stimulus in the natural pacemaker node.

The expected normal heart rhythm is presented in Figure~\ref{fig:normal} showing a close agreement between the real ECG signal and the simulated one, respectively Figure~\ref{fig:normal}(a) and Figure~\ref{fig:normal}(b). It should be pointed out that simulations capture the main features of the real ECG signal, characterized by P, QRS and T waves. The Figure~\ref{fig:normal}(c)-(e) show the individual components of the ECG signal related to each node.

The pathological behaviors shown in Figure~\ref{fig:diseases} are obtained changing the amplitude and frequency of the external oscillator in the SA node. Only four combinations of these parameters were sufficient to identify in the ECG some patterns that can indicate diseases. The double R peaks, for example, are associated with branch blocks \cite{canabrava2014eletrocardiografia}, which are related to delays in the transmission of the electrical impulses in the heart. The alternation of T waves are very useful to indicate cardiac sudden death \cite{barbosa2004alternancia}. In the case of P wave alternans, is possible infer junctional tachycardia \cite{brugada1991new}, a form of tachycardia with the involvement of the AV node. The last pattern shown in the ECGs is the ventricular tachycardia. This last pathological rhythm is associated with high-frequency ventricular contraction. 

Another point to be highlighted is regarding the cardiac frequency and its distribution. Differently of the normal rhythm, which the heart rate is possible to maintain constant in 90 bpm, where the normal behavior can range from 60 to 100 bpm \cite{haddad2006evolution}, the pathology presented in the Figure~\ref{fig:diseases}(c), considering 30 minutes of simulation, show us the heart operating in a high-frequency rate and highly dispersed.

\begin{figure}[htb]
    \centering
    \includegraphics[width=\textwidth]{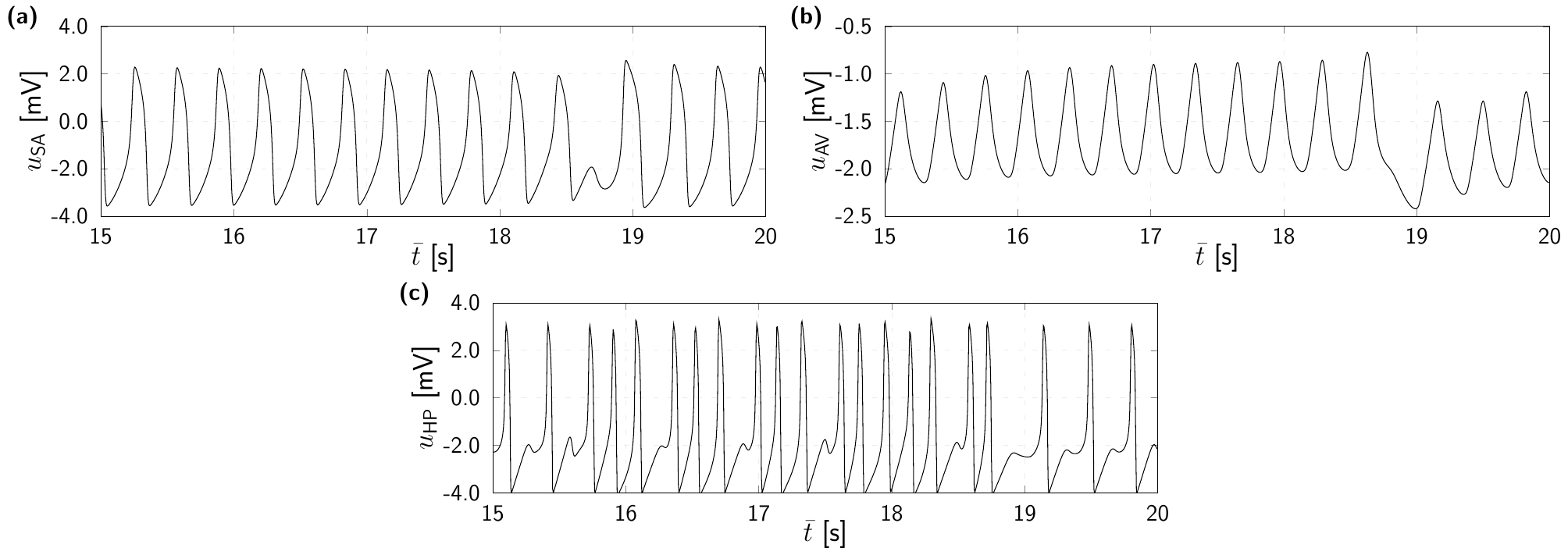}
    \caption{Components of the simulated ECG of the pathological behavior for $\rho_\text{SA} = 9.6$ and $\omega_\text{SA} = 2.1$: (a) SA node; (b) AV node; (c) HP node.}
    \label{fig:disease1}
\end{figure}

Considering, for instance, the external excitation parameters being $\rho_\text{SA} = 9.6$ and $\omega_\text{SA} = 2.1$, the Figure~\ref{fig:disease1} shows the correspondent components of each node. As consequence of the external stimulus, the oscillation frequency of the SA signal is increased and, due to the coupling terms, both AV and HP signals are dramatically changed when compared to the normal behavior. The combination of these signals produces the pathological rhythm as depicted in the ECG signal in the Figure~\ref{fig:diseases}.

Now, in order to turn these pathological rhythms into normal ones, the proposed controller is introduced in the next section.

\section{Sliding Mode Control with Gaussian Process Regression}

In this section, the sliding mode controller for the natural pacemaker is presented. Then, the Gaussian process regressor is shown as compensator of the modelling inaccuracies and unmodeled dynamics. After that, the Lyapunov stability analysis is called to demonstrate the boundedness and convergence properties of the controller.

\subsection{Sliding Mode Control for the Natural Pacemaker}

Heart dynamics is a spatiotemporal and multiphysics phenomenon, but the corresponding electrical activity can be described by nonlinear delayed differential equations (DDEs), which constitute a reduced-order model for the description of cardiac rhythms. For the SA node, the numerical model \eqref{eq:cheffer_savi1} can be rewritten as follows:
\begin{equation}\label{eq:general}
    \ddot{u}_{\text{SA}} = f_{\text{SA}}
\end{equation}
where $f_{\text{SA}}$ represents the vector field corresponding to equation~\eqref{eq:cheffer_savi1} and $(u_{\text{SA}},\dot{u}_{\text{SA}})$ are the states to be controlled.

However, it can be very useful for different purposes, such as the design of control schemes, a simplified version of the equation \eqref{eq:general} that highlights the unmodeled dynamics and model inaccuracies. Therefore, in view of the design of a control system for the natural pacemaker, the numerical model of the SA node will be rewritten here in the following form:
\begin{equation}\label{eq:general_c}
    \ddot{u} = \hat{f} + v + d
\end{equation}
where the the SA index is omitted to simplify the notation and which $\hat{f}$ stands for what is known by the designer about $f$, $v$ is the control signal, assumed to be applied to the SA node, and $d$ takes account of the unknown dynamics, parametric uncertainties and also the occasional perturbations.

Now, according to the sliding mode method \cite{slotine1991applied}, a sliding surface can be defined as follows:
\begin{equation}\label{eq:s}
    s = \dot{\tilde{u}} + \lambda \tilde{u}
\end{equation}
with $\tilde{u} = u - u_\text{d}$ representing the tracking error associated with the desired state $u_\text{d}$ and $\lambda$ being a strictly positive constant. In order to avoid the undesirable chattering effect, a saturation function is proposed to replace the standard discontinuity term $\sgn(\cdot)$ and to represent the boundary layer neighboring the sliding surface \cite{slotine1991applied}:
\begin{equation}\label{eq:sat}
    \sat (u) = \left\{\begin{array}{ccc}
      u   & \text{if} & |u| \leq 1,  \\
      \sgn (u) & \text{if} & |u| > 1.
    \end{array}\right.
\end{equation}

Therefore, the smooth sliding mode control law for a system represented by equation~\eqref{eq:general_c} can be designed as follows:
\begin{equation}\label{eq:law}
    v = -\hat{f} - \hat{d} + \ddot{u}_\text{d} - \lambda \dot{\tilde{u}} - \kappa \sat (s/\phi)
\end{equation}
with $\hat{d}$ being, respectively, estimate of $d$, $\kappa$ representing the control gain, and $\phi$ being a strictly positive constant denoting the width of the boundary layer.

Applying the control law~\eqref{eq:law} to~\eqref{eq:general_c} and assuming that the closed-loop states are inside the boundary layer when $\vert s \vert \leq \phi$, we get
\begin{equation}\label{eq:closed-loop}
    \dot{s} + \kappa/\phi s = \delta
\end{equation}
with $\delta = d - \hat{d}$ being the approximation error. From \eqref{eq:closed-loop} it can be seen that in the case of perfect estimation, i.e.\ $\hat{d}=d$, the combined error $s$ and therefore the tracking error $\tilde{u}$ converges to zero. Otherwise, closed-loop dynamics is driven by the approximation error $\delta$. Furthermore, it suggests that the signal $s$ may also represent a reasonable metric and help the GP regressor to compute the estimate $\hat{d}$.

\subsection{Gaussian Process Regression}

Considering that Gaussian process regression can defined as nonparametric model to describe a distribution over functions, the estimate of the total uncertainty $d$ will be computed as a function of the sliding surface $s$ with noisy observations \cite{williams2006gaussian} as follows:
\begin{equation}
    \bar{d} = d(s) + \varepsilon, \quad \varepsilon \sim \mathcal{N}(0, \sigma_\varepsilon^2)
\end{equation}
where $\varepsilon$ é the noisy signal obtained by a normal distribution with null mean and variance $\sigma_\varepsilon^2$.

The Gaussian process can be understood as a distribution of probabilities over the function space of $d$, since that any finite set of function values is jointly Gaussian \cite{williams2006gaussian}. The GP distribution is defined by a mean $\mu(s) = \mathbb{E}[d(s)]$ and a covariance $k(s, s^\prime) = \mathbb{V}[d(s), d(s^\prime)]$ functions:
\begin{equation}
    d(s) \sim \mathcal{GP}[d(s), k(s, s^\prime)]
\end{equation}

Given a training set $\mathcal{D}_N = \{ s_i, d_i \}_{i=1}^N$, the GPR learns a function by a prediction of the (normal) distribution of functions values $d(s^\ast)$ at arbitrary inputs $s^\ast$ by the following posterior mean and variance functions:
\begin{subequations}
\begin{equation}
    \mathbb{E}[d(s^\ast \vert \mathcal{D}_N)] = \mu(s^\ast) + k(s^\ast, s)^\top (\bm{K}_N + \sigma_\varepsilon^2 \bm{I})^{-1} \tilde{d}(s)
\end{equation}
\begin{equation}
    \mathbb{V}[d(s^\ast \vert \mathcal{D}_N)] = k(s^\ast, s^\ast) - k(s^\ast, s)^\top (\bm{K}_N + \sigma_\varepsilon^2 \bm{I})^{-1} k(s^\ast, s)
\end{equation}
\end{subequations}
which $\bm{K}_N$ stands for the covariance matrix defined as $K_{i,j} = k(s_i, s_j)$ and $\tilde{d}(s) = [\bar{d}(s_1) - \mu(s_1) ~\ldots~ \bar{d}(s_N) - \mu(s_N)]^\top$.

It is proposed here that the disturbance compensation be calculated by the predictive mean function, i.e. $\hat{d} = \mathbb{E}[d(s^\ast \vert \mathcal{D}_N)]$. On the other hand, the variance function works here to delimit the bounds of $d$. For this, it will be assumed that the disturbance term is limited by the posterior variance in the form $\hat{d} - \vartheta \sigma \leq d \leq \hat{d} + \vartheta \sigma$, with $\vartheta$ being the confidence level and $\sigma^2 = \mathbb{V}[d(s^\ast \vert \mathcal{D}_N)]$. This confidence interval is essential to design the control gain by the stability analysis on the next subsection.

Regarding the GPR algorithm, it is important to highlight the rolling window process about the generation of the training set $\mathcal{D}_N$. Initially, the training set is empty. Given the entries $s$, computed by the equation \eqref{eq:s}, and $\bar{d} = \ddot{u} - \hat{f} - v$ from the equation \eqref{eq:general_c}, these variables are added to the budget. When the GP dictionary maximum size is achieved, the oldest ones are discarded to allow the addition of new entries. We emphasize that the maximum length is necessary in order not to exaggerate the computational cost of implementation.

\subsection{Boundedness and Convergence Analysis}

The boundedness and convergence properties of the closed-loop signals in the presence of modeling inaccuracies can be investigated by means of a Lyapunov stability analysis. Thus, let a positive-definite function $V$ be defined as:
\begin{equation}\label{eq:lyap}
    V(t) = \frac{1}{2} s_\phi^2
\end{equation}
where $s_\phi = s - \phi\sat(s/\phi)$ is the distance between $s$ and the boundary layer. 

Since $\dot{s}_\phi = \dot{s}$ outside the boundary layer and applying the equations \eqref{eq:general_c} and \eqref{eq:law}, the time derivative of $V$ becomes:
\begin{equation*}
\begin{split}
    \dot{V}(t) & =  s_\phi \dot{s}\\
               & =  s_\phi[\ddot{u} - \ddot{u}_d + \lambda \dot{\tilde{u}}] \\
               & =  s_\phi[\hat{f} + v + d - \ddot{u}_d + \lambda \dot{\tilde{u}}] \\
               & = s_\phi[d - \hat{d} - \kappa \sat (s/\phi)]
\end{split}
\end{equation*}

Since inside the boundary layer $s_\phi = 0$, we get $\dot{V}(t) = 0$, outside of that region $\sat(s/\phi) = \sgn(s_\phi)$, which implies that $\dot{V}$ becomes:
\begin{equation}\label{eq:lyapdot}
    \dot{V}(t) = -[\hat{d} - d + \kappa \sgn (s_\phi) ]s_\phi
\end{equation}

Then, it follows from the bounds of $d$ that the confidence interval can be a tuning function of the control gain $\kappa$ by the below consideration:
\begin{equation}\label{eq:cont_gain}
    \kappa > \eta + \vartheta \sigma
\end{equation}
where $\eta$ is a strictly positive parameter. Therefore, applying the condition \eqref{eq:cont_gain} in \eqref{eq:lyapdot}:
\begin{equation}
    \dot{V}(t) \leq -\eta \vert s_\phi \vert
\end{equation}
which implies $V(t) \leq V(0)$ and that any initial state will be attracted to the boundary layer. Moreover, remembering that $V(t)$ is positive definite, its time derivative $\dot{V}(t)$ is equal to zero inside the boundary layer and $\dot{V}(t) < 0$ outside of it, so the states will remain in the boundary layer as $t \rightarrow \infty$.

Once inside the boundary layer, the closed-loop states will be bounded in a region around the origin. Then, it follows that $\vert s \vert \leq \phi$. Hence, remembering that $s = \dot{\tilde{u}} + \lambda \tilde{u}$, we have
\begin{equation}\label{eq:bounds1}
    -\phi \leq \dot{\tilde{u}} + \lambda \tilde{u} \leq \phi
\end{equation}

Thus, multiplying \eqref{eq:bounds1} by $e^{\lambda t}$ gives
\begin{equation}\label{eq:bounds2}
    -\phi\,e^{\lambda t} \leq \frac{d}{dt} \left( \tilde{u} e^{\lambda t} \right) \leq \phi\,e^{\lambda t}
\end{equation}

Integrating \eqref{eq:bounds2} between 0 and $t$ yields
\begin{equation}\label{eq:bounds3}
    -\frac{\phi}{\lambda} e^{\lambda t} -  \left[ \vert\tilde{u} (0)\vert + \frac{\phi}{\lambda} \right] \leq \tilde{u} e^{\lambda t} \leq \frac{\phi}{\lambda} e^{\lambda t} + \left[ \vert\tilde{u} (0)\vert + \frac{\phi}{\lambda} \right]
\end{equation}

By dividing \eqref{eq:bounds3} by $e^{\lambda t}$
\begin{equation}\label{eq:bounds4}
    -\frac{\phi}{\lambda} -  \left[ \vert\tilde{u} (0)\vert + \frac{\phi}{\lambda} \right] e^{- \lambda t} \leq \tilde{u} \leq \frac{\phi}{\lambda} + \left[ \vert\tilde{u} (0)\vert + \frac{\phi}{\lambda} \right]  e^{- \lambda t}
\end{equation}
it follows, for $t \rightarrow \infty$, that
\begin{equation}\label{eq:bounds5}
    -\frac{\phi}{\lambda}  \leq \tilde{u} \leq \frac{\phi}{\lambda}
\end{equation}

Applying \eqref{eq:bounds5} to \eqref{eq:bounds1}, it can be verified that
\begin{equation}\label{eq:bound2}
    -2\phi  \leq \dot{\tilde{u}} \leq 2\phi
\end{equation}

Therefore, it is possible to conclude that the controller ensures the exponential convergence of the tracking error to the closed region $\mathcal{U} = \{ (\tilde{u},\dot{\tilde{u}}) \in \mathbb{R}^2 : \vert\tilde{u}\vert \leq \phi/\lambda \: \text{and} \: \vert\dot{\tilde{u}}\vert \leq 2\phi\}$.

\section{Rhythm Control}

The proposed controller is now evaluated by means of numerical simulations at a sampling rate of 100 Hz. The desired states are extracted from a expected normal heart cycle for the natural pacemaker, which means that the controller's main goal is to achieve a normal rhythm in the ECG controlling the SA node behavior while avoiding cardiac disorders. On this basis, the pathology investigated is related the most critical shown in the figure~\ref{fig:diseases}: $\rho_\text{SA} = 9.6$ and $\omega_\text{SA} = 2.1$.

The controller's parameters are set to $\lambda = 3$, $\phi = 20$, $\eta = 3$, and $\vartheta = 2$. Assuming that no prior knowledge about the cardiac system is available to the control system designer, i.e.\ $\hat{f} = 0$, the ability of $\hat{d}$ to handle all neglected dynamical effects is investigated. The Gaussian process regressor is implemented to compute $\hat{d}$. In this task, it is assumed that the kernel function is of squared exponential type:
\begin{equation}\label{eq:kernel}
    k(s, s^\prime) = \sigma_f^2 \exp \left[ -\frac{(s - s^\prime)^2}{2 \ell^2} \right]
\end{equation}
where $\sigma_f$ is the prior standard deviation and $\ell$ the length-scale. These parameters with the noise standard deviation $\sigma_\varepsilon$ represent the hyperparameters of the GP distribution. In the control simulation, it is assumed that the prior mean is set to zero while the hyperparameters are defined as $\sigma_f = 1$, $\ell = 10^{-6}$, and $\sigma_\varepsilon = 0.2$. The rolling window slides with the training set size maintained constant at $N = 50$.

\begin{figure}[htb]
    \centering
    \includegraphics[width=\textwidth]{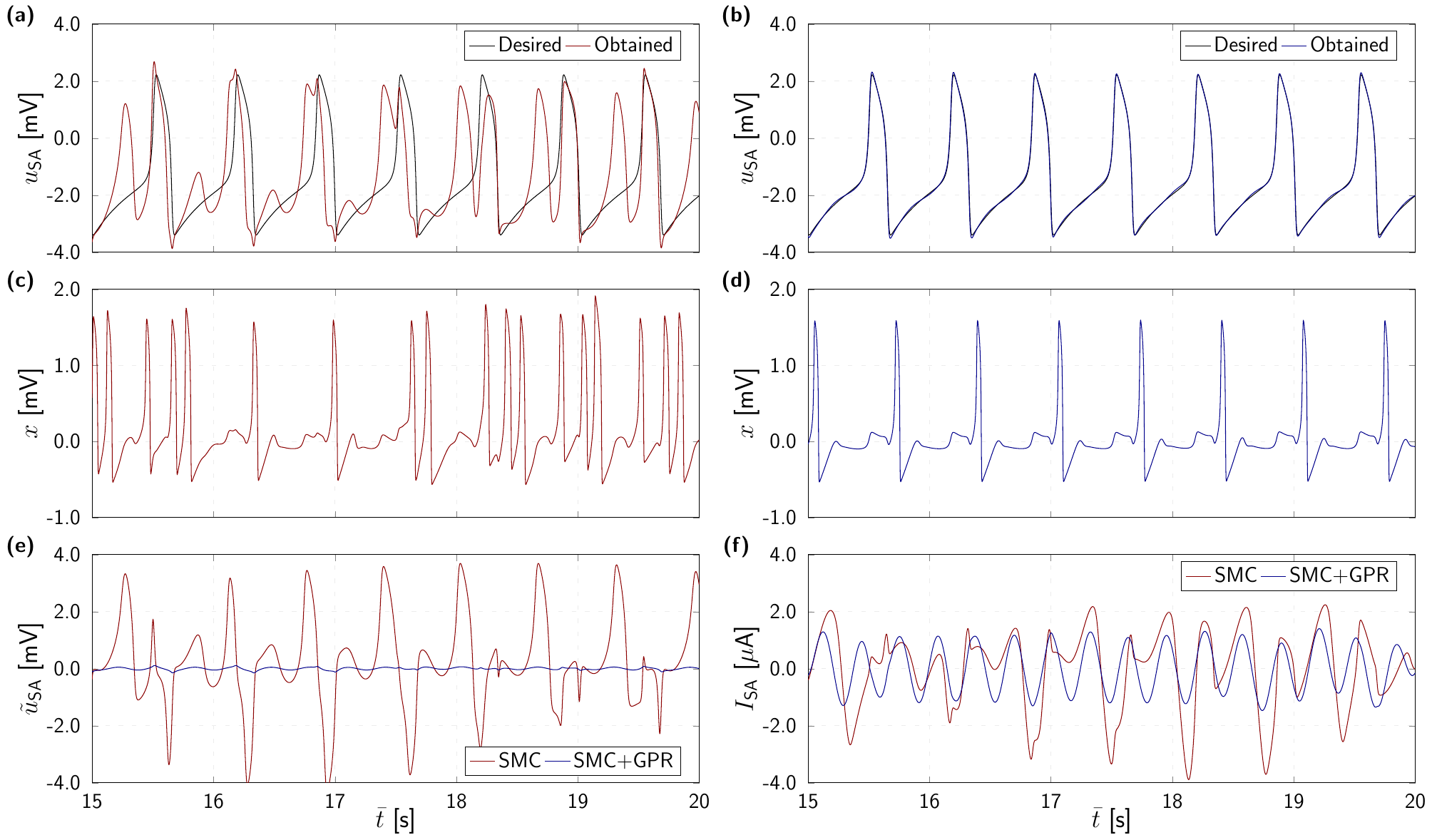}
    \caption{Simulation results: (a) SA node control with the smoothed SMC, (b) SA node control with the proposed controller (SMC+GPR), (c) resulting ECG with the smoothed SMC, (d) resulting ECG with the proposed controller, (e) control error, and (f) control signal.}
    \label{fig:results}
\end{figure}

\begin{figure}[htb]
    \centering
    \includegraphics[width=0.5\textwidth]{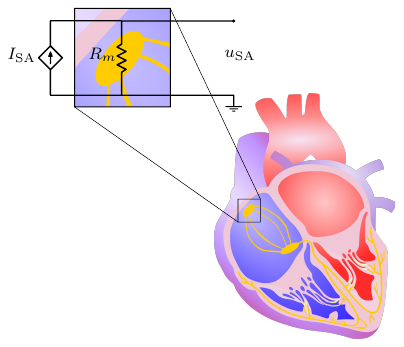}
    \caption{Framework of a artificial pacemaker circuit.}
    \label{fig:pacemaker}
\end{figure}

Figure~\ref{fig:results} shows a comparison between conventional and proposed control schemes applied to the control in the natural pacemaker. As can be seen, the proposed approach is able to stabilize the expected normal rhythm in the SA node, Figure~\ref{fig:results}(b), while the conventional one fails, Figure~\ref{fig:results}(a). As consequence of the normalization of the SA signal, the presented controller, Figure~\ref{fig:results}(d), regulates the ECG signal converting in the a normal rhythm with clearly identification of the P, QRS and T waves and decreasing the heart rate to around 90 bpm, while with the smoothed controller, Figure~\ref{fig:results}(c), in the ECG can be observed double R peaks, T wave alternans, and ventricular tachycardia, which indicates abnormal functioning of the heart. It should also be noted that the proposed scheme drastically reduces the control error, Figure~\ref{fig:results}(e), when compared with the conventional scheme. In order to make the application of the proposed controller more realistic, the control signal $v$, Figure~\ref{fig:results}(f), was converted in electric current. For this purpose, it was considered that a artificial pacemaker, modeled as a variable current source, can be implemented directly in the SA node, as illustrated in the Figure~\ref{fig:pacemaker}. The current $I_\text{SA}$ is calculated dividing the double integration of $v$ by the membrane resistance of the SA node for an average adult, $R_m = 20~\Omega$ \cite{grant1982intracellular}.

\begin{figure}[htb]
    \centering
    \includegraphics[width=0.5\textwidth]{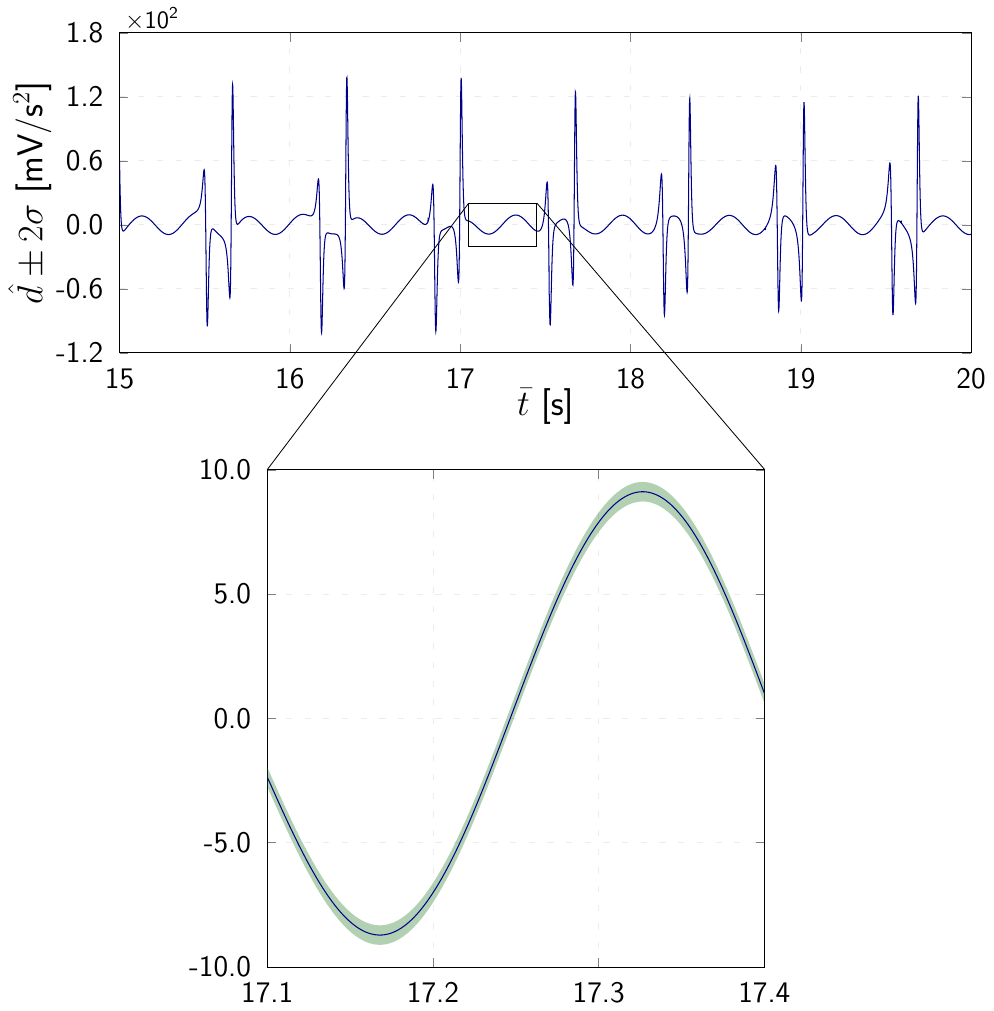}
    \caption{Gaussian process estimation of the unknown dynamics with its uncertainty bounds.}
    \label{fig:results_gp}
\end{figure}

The success of the proposed controller is due the ability of the Gaussian process regressor in predict and compensate the unknown dynamics. The figure \ref{fig:results_gp} shows the GPR compensator $\hat{d}$ with its uncertainty bounds $\pm 2 \sigma$ used to compute the control gain. In fact, while the predictive mean is fundamental to compensate the disturbance and modeling inaccuracies which implies a reduced level of tracking error when compared to the conventional control strategy, the predictive standard deviation plays an essential role in providing the proper bounds of $\hat{d}$ decreasing the control gain and, consequently, the control effort level due to the proportionality between $\kappa$ and $\sigma$, and ensuring trajectory tracking even when the regressor is not able to reproduce $d$.

\section{Concluding remarks}\label{sec13}

This paper investigates the control of the electrical activity of the heart from the natural pacemaker's point of view. The sliding mode approach works as the main framework of the proposed controller and a supervised machine learning technique based on Gaussian processes as unknown dynamics compensator. A mathematical model consisting of three coupled nonlinear oscillators is employed to describe cardiac rhythms, being able to represent both normal and pathological behaviors. The boundedness and convergence properties of the SA node closed-loop signals are analytically proved by means a Lyapunov stability analysis. The controller's effectiveness is evaluated by numerical simulations whose comments of the obtained results are summarized as follows:

\begin{enumerate}
    \item The proposed controller is capable to deal with the nonlinear nature of the heart dynamics, allowing accurate trajectory tracking of the normal cardiac rhythm at the SA node;
    \item The Gaussian process regression can learn, predict, and compensate the modeling inaccuracies and disturbances improving the controller performance;
    \item An online and adaptive approach, rather than offline procedure, can be properly adopted to allow a continuously learning about the dynamical changes in the heart behavior;
    \item These features allow the adopted Gaussian regressor to adapt to different individuals and continuously approach their cardiac dynamics, which confers the ability to deal with intra and interpatient variability.
    \item The use of one single signal as input as well as the rolling windows scheme with limitation in the training set size allows a considerable reduction of the complexity of computational implementation, making possible its application in low-power circuits of artificial pacemakers;
    \item A simple representation of an electric circuit can be used to understand the control signal conversion and, consequently, its amplitude scale in terms of electric current.
\end{enumerate}

Therefore, it can be verified that the sliding mode controller embedded with a Gaussian process regressor can handle successfully with heart rhythm normalization, even with a high level of uncertainty about the system dynamics, avoiding critical cardiac behaviors. From the results presented, the experimental implementation is a promising step towards the final validation of the proposed controller.

\bibliographystyle{plain}
\bibliography{heart}

\end{document}